\documentclass[journal,twocolumn,web]{ieeecolor}
\usepackage{cite}
\usepackage{amsmath,amssymb,amsfonts}
\usepackage{algorithmic}
\usepackage[per-mode = symbol, free-standing-units,unit-optional-argument,space-before-unit,use-xspace]{siunitx}
\usepackage{graphicx}
\usepackage{generic}
\usepackage{algorithm,algorithmic}
\usepackage{hyperref}
\hypersetup{hidelinks}
\usepackage{comment}
\usepackage{multirow}

\newcommand{\shortit}{}
\newcommand{\longit}{\textit}

\DeclareSIUnit\cm{\centi\meter}
\newcommand{\val}[2][\centi\meter]{\qtylist{#2}{#1}} 

\definecolor{subsectioncolor}{rgb}{0,0,0} 

\usepackage{etoolbox}
\makeatletter
\@ifundefined{color@begingroup}%
  {\let\color@begingroup\relax
   \let\color@endgroup\relax}{}%
\def\fix@ieeecolor@hbox#1{%
  \hbox{\color@begingroup#1\color@endgroup}}
\patchcmd\@makecaption{\hbox}{\fix@ieeecolor@hbox}{}{\FAILED}
\patchcmd\@makecaption{\hbox}{\fix@ieeecolor@hbox}{}{\FAILED}

\usepackage{textcomp}
\def\BibTeX{{\rm B\kern-.05em{\sc i\kern-.025em b}\kern-.08em
    T\kern-.1667em\lower.7ex\hbox{E}\kern-.125emX}}
\usepackage{dirtytalk}

\begin{document}
\bstctlcite{Refs:use_etal} 

\title{LuViRA Dataset Validation and Discussion: Comparing Vision, Radio, and Audio Sensors for Indoor Localization}
\author{Ilayda Yaman, Guoda Tian, Erik Tegler, Jens Gulin, Nikhil Challa, Fredrik Tufvesson, \\
Ove Edfors, Kalle Åström, Steffen Malkowsky, and Liang Liu
\thanks{This work is funded by the Swedish Research Council and Ericsson AB and partially supported by the Wallenberg AI, Autonomous Systems and Software Program (WASP) funded by the Knut and Alice Wallenberg Foundation.
\textit{(Corresponding author: Ilayda Yaman.)}} 
\thanks{The authors are all with LTH, Lund University, SE-22100~Lund, Sweden (Corresponding author email: ilayda.yaman@eit.lth.se).}}

\maketitle

\begin{abstract}
We present a unique comparative analysis, and evaluation of vision, radio, and audio based localization algorithms.
We create the first baseline for the aforementioned sensors using the recently published Lund University Vision, Radio, and Audio (LuViRA) dataset, where all the sensors are synchronized and measured in the same environment. 
Some of the challenges of using each specific sensor for indoor localization tasks are highlighted. Each sensor is paired with a current state-of-the-art localization algorithm and evaluated for different aspects: localization accuracy, reliability and sensitivity to environment changes, calibration requirements, and potential system complexity. Specifically, the evaluation covers the ORB-SLAM3 algorithm for vision-based localization with an RGB-D camera, a machine-learning algorithm for radio-based localization with massive MIMO technology, and the SFS2 algorithm for audio-based localization with distributed microphones. 
The results can serve as a guideline and basis for further development of robust and high-precision multi-sensory localization systems, e.g., through sensor fusion, context, and environment-aware adaptation.  
    \end{abstract}
    

\section{Introduction}
\label{sec:introduction}
\IEEEPARstart{T}{he} need for accurate and reliable location information is rising with the increasing number of applications that require seamless localization, such as smart factories, autonomous driving, and unmanned aerial vehicles (UAVs)~\cite{seamless_loc}. 
Achieving seamless localization is a critical prerequisite for ensuring the safety, efficiency, and reliability of operations in many scenarios, especially in dynamic conditions such as urban mobility, indoor navigation, and rapidly changing environments. 
For outdoor scenarios, the real-time kinematic enabled global navigation satellite system (GNSS-RTK) can achieve cm-level accuracy under good conditions~\cite{gnss}. For indoor scenarios, there exist many different systems using various sensors and technologies for performing localization tasks, including WiFi, Bluetooth, digital cameras, inertial measurement units (IMU), and microphones. Different sensors have their advantages and limitations and there is a need for an in-depth investigation in this area. This investigation requires vision, radio, and audio sensor data to be captured in the same environment at the same time. However, there is very limited work that evaluates these sensors jointly with a realistic dataset, while using 5G technologies such as massive multiple-input multiple-output (MIMO). 

Our paper conducts a detailed evaluation of selected existing vision, radio, and audio based localization algorithms and compares their localization performances in a real indoor scenario. To this end, our main contributions are as follows:

\begin{itemize}
    \item  Providing baseline results for the first publicly available dataset that contains synchronized data from vision, radio, and audio sensors, and a ground truth system with localization error less than $0.5$\mm: the Lund University Vision, Radio, and Audio (LuViRA) dataset~\cite{LuViRA}\footnote{ \url{https://github.com/ilaydayaman/LuViRA_Dataset}}.  
    \item  Evaluating each sensor type for localizing the target object based on the following aspects using state-of-the-art localization algorithms: localization accuracy, reliability and sensitivity to environment changes, calibration requirements, and potential system complexity. 
    \item  Analyzing the advantages and limitations of localization algorithms based on aforementioned sensors coexisting in the same environment for trajectories with static and dynamic (people or objects moving around, sharp movements) environments. 
\end{itemize}
 
\begin{figure}[tb!]
  \centering
  \includegraphics[width=1.0\linewidth]{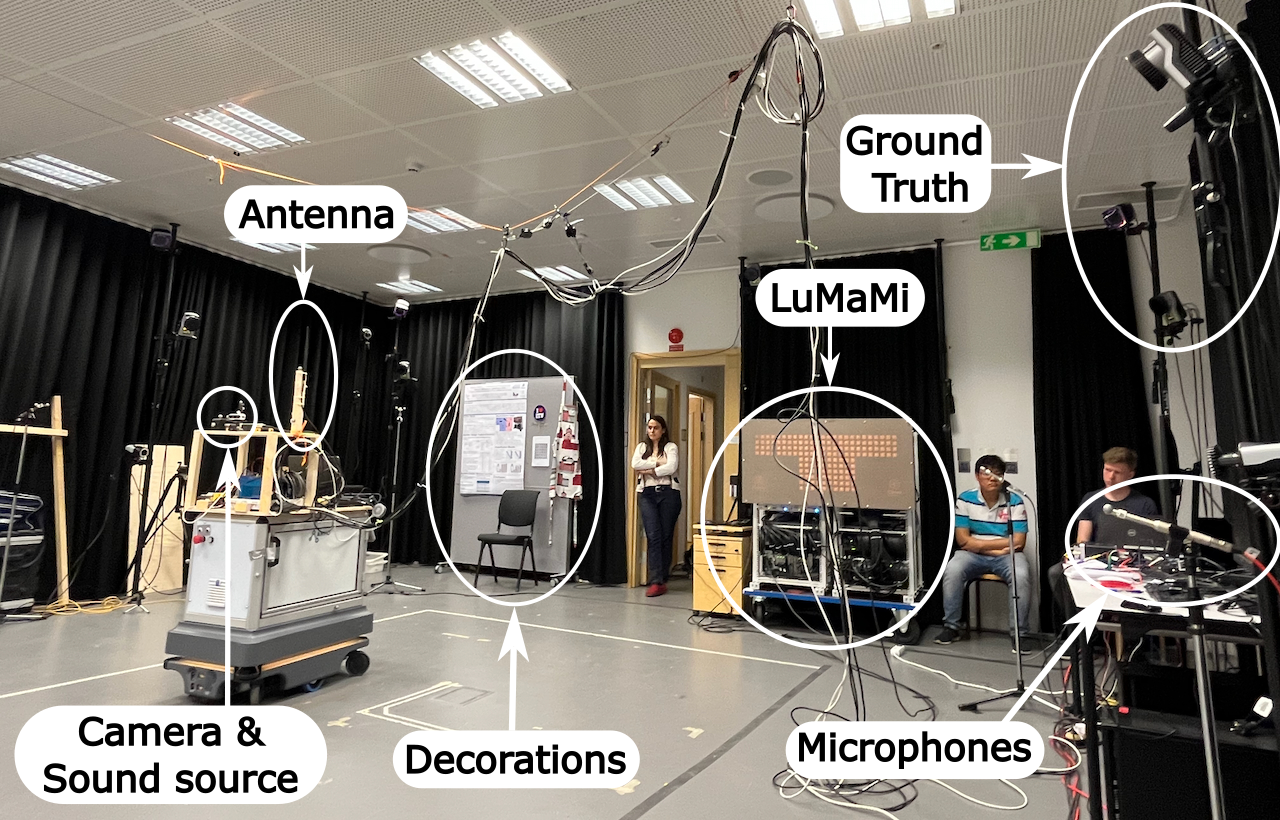}
  \caption{An overview of the measurement setup during the recording of the \longit{Random circle1} trajectory.}
  \label{fig:Measurements} 
\end{figure}

\noindent The LuViRA dataset contains massive MIMO channel data captured using the $100$-antenna \textit{Lund University massive MIMO} (LuMaMi) testbed~\cite{LUMAMI}, RGB-D image streams from an Intel® RealSense depth camera D435i, and audio clips captured by $12$ microphones. An overview of the measurement environment can be seen in Fig.~\ref{fig:Measurements}. We use the $89$ trajectories provided in the LuViRA dataset that contains all the sensors with different movement patterns for performance evaluation.

The state-of-the-art algorithms chosen to evaluate the localization accuracy of each sensor are \textit{Oriented FAST and Rotated BRIEF simultaneous localization and mapping} (ORB-SLAM3)~\cite{ORB-SLAM3}, a high precision machine-learning (ML) algorithm~\cite{ICC}, and
StructureFromSound2 (SFS2)~\cite{9616051}\footnote{ \url{https://github.com/kalleastrom/StructureFromSound2}}. Based on the difference between the estimated location from these algorithms and the ground truth labels included in the dataset, the localization accuracy is calculated in terms of the mean and standard deviation (SD) of the localization errors. To assess the robustness of the algorithms and sensors, the localization accuracy is evaluated under different conditions, e.g., static and dynamic environments. The sensitivity to calibration errors of different sensors and algorithms is also discussed, from the general aspects of required calibration to precision and frequency requirements. \textbf{To the best of our knowledge, this is the first paper that evaluates and compares localization algorithms based on vision, 5G radio, and audio sensors in a realistic and synchronized measurement scenario.}

\section{Related Works}
Many academic works on indoor localization study vision, radio, and audio sensors individually. For example, there are many localization systems that use variations of visual sensors such as monocular and stereo, or RGB-D cameras~\cite{ORB-SLAM3}\cite{Kendall2015PoseNetAC}. Moreover, the camera information is frequently fused with other sensors such as IMU, and Light Detection and Ranging (Lidar), to combat the limitations of the vision-based localization systems in dark, low-texture, and dynamic environments~\cite{rovio}. 
On the other hand, the localization accuracy of traditional radio-based localization algorithms, such as Bluetooth and Wi-fi based fingerprinting, are highly dependent on the setup and measurement environment. The accuracy requirements are increasing with each generation and extremely high accuracy is expected to be fundamental for the sixth generation (6G) systems~\cite{6g_localization}. Thus, there are many works that explore new enabling technologies, such as distributed large antenna arrays and millimeter-wave transmissions, to improve localization accuracy and robustness~\cite{jesus_algorithm}. 
One enabler is massive MIMO antenna arrays, that provide high spatial resolution to fulfill the requirements of accurate localization~\cite{mimo_dataset}\cite{joaos_algorithm}. 
Even though audio-based localization is more of a niche area compared to vision and radio based approaches, various methods are available for localizing sound sources~\cite{TDOA_1}\cite{zhayida2014automatic}. Under-water localization, such as passive sonar, relies on audio signals since it has good propagation properties where both vision and radio are weak in performance~\cite{Su2020ARO}.
In many cases, the object to localize naturally has an audio signature, whereas vision relies on a separate light source and radio requires active transmission. 
There are very few works that use vision, radio, and audio sensors jointly to improve the accuracy, reliability, and robustness of the localization task~\cite{radio-visual}\cite{Audio_Visual}. When there are discussions on these sensors together, the sensors are discussed in different scenarios~\cite{survey_indoor}. 
Moreover, none of these evaluations are based on a public dataset that includes synchronized data of all the aforementioned sensors from the same environment.  
We believe the number of works that fuse these sensors will increase with the aid of a public dataset with known baselines, and an extensive study on the advantages and limitations of each sensor.  
 

\section{Background} 
In this section, the localization algorithms selected for different sensor systems are presented at a high level of abstraction, to provide necessary background information.

\subsection{Vision System}

Most vision-based localization algorithms assume no prior knowledge of the environment. Thus the problem of localization extends to simultaneous localization and mapping (SLAM). Briefly, a classical Visual SLAM framework consists of the following steps~\cite{gao2021introduction}: 
\begin{enumerate}
   \item Feature extraction detects and matches key points in each frame.
   \item Visual odometry (VO) creates an initial local map by estimating the movement, location, and orientation (pose) of the camera from consecutive frames. 
   \item Loop closing detects if the camera goes back to a previous location.
   \item Filtering and optimization (i.e. bundle adjustment) create a more accurate camera trajectory and an environment map using estimates from VO and loop closing. 
\end{enumerate}

\begin{figure*}[tb!]
  \centering
  \includegraphics[width=0.9\linewidth]{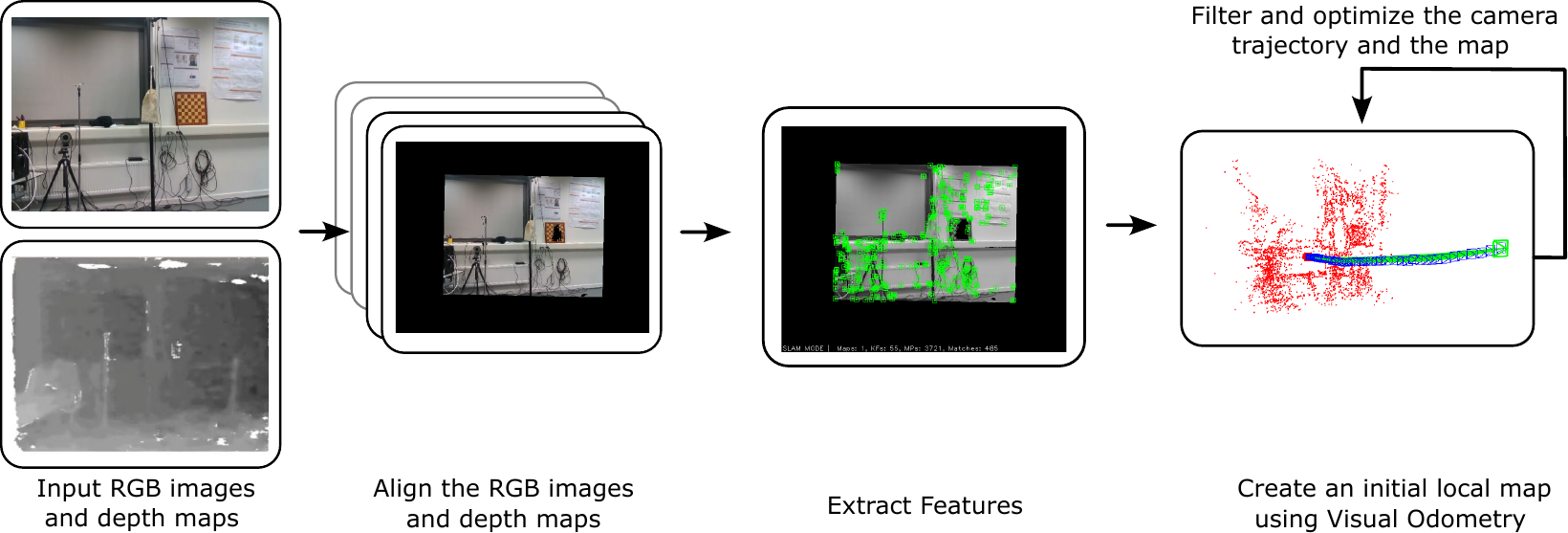}
  \caption{Selected vision-based localization algorithm pipeline.}
  \label{fig:block_diagram_vision} 
\end{figure*}

Popular algorithms that solve the SLAM problem include ORB-SLAM3~\cite{ORB-SLAM3}, ROVIO~\cite{rovio}, and RTAB-MAP~\cite{RTABMapAA}. 
ORB-SLAM3 is a very well-established, open-source algorithm. Oriented FAST and Rotated BRIEF (ORB) features are used in the visual odometry and loop closing steps, resulting in accurate, robust, and real-time localization and mapping, even in dynamic environments. 
Moreover, ORB-SLAM3 is designed to support different types of cameras depending on the available hardware in the system. For example, if there is only a single camera available, the monocular ORB-SLAM3 algorithm is used for localization. The accuracy of the monocular ORB-SLAM3 is however limited due to the lack of depth information on the extracted features. Instead, if there is a stereo or depth camera in the environment, a more robust and accurate solution can be obtained by using stereo or RGB-D ORB-SLAM3 algorithms, which are used in this paper for vision-based localization. A high-level diagram of the pipeline using the LuViRA dataset is shown in Fig. \ref{fig:block_diagram_vision}. The IMU data from the camera unit is not used.

\subsection{Radio System}
The accuracy and applicability of different radio-based localization systems differ greatly from each other depending on the system configuration, such as system bandwidth, number of antennas, or input features used. The following three features are widely used:
\begin{itemize}
    \item \textbf{Received signal strength (RSS)}: The distance between the transmitter and receiver can be estimated by using a propagation path loss model. This model establishes a link between the received signal strength and the separation distance between the transmitter and receiver. 
    \item \textbf{Delay or time based:} The most common time-based methods use the time of arrival (TOA) and time difference of arrival (TDOA), which generally require multiple synchronized receivers. TOA calculates the distance between the transmitter and the receiver by multiplying the signal propagation time with the speed of light. TDOA computes the arrival time differences at the receivers. The location of the transmitter can be estimated from the intersections of hyperboloids created by those time differences. 
    \item \textbf{Angle of Arrival (AOA):} Arrival angles of the received signals are calculated and used to estimate the location of the transmitter by triangulation or multilateration techniques using either multiple AOAs or a single AOA together with the measured TOA. 
\end{itemize}

\begin{figure*}[tb!]
  \centering
  \includegraphics[width=0.70\linewidth]{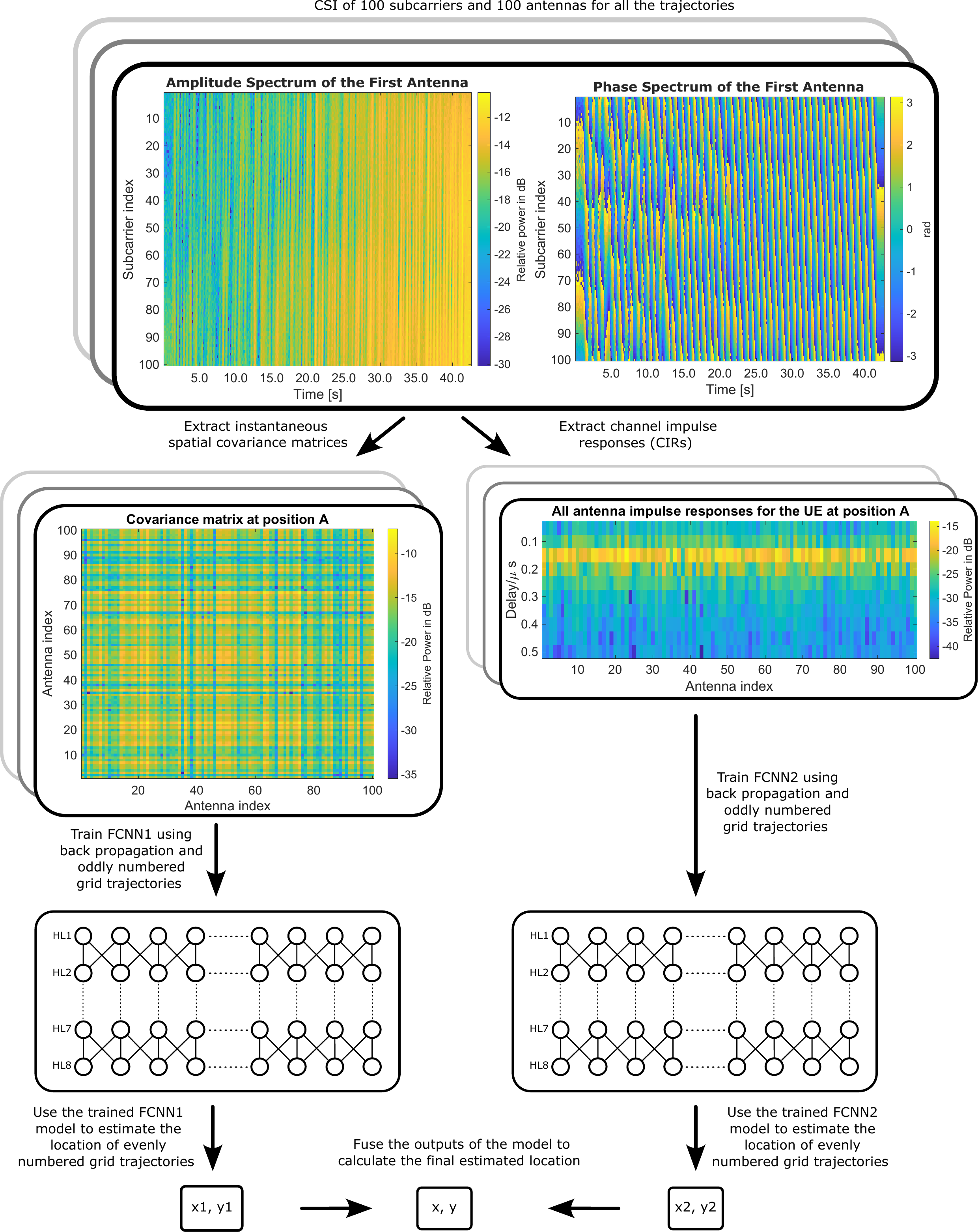}
  \caption{Selected radio-based localization algorithm pipeline where the fully connected neural network layer 1 (FCNN1) extracts angular information and FCNN2 extracts delay information through the respective hidden layers (HL).}
  \label{fig:block_diagram_radio} 
\end{figure*}

\noindent Time and angle based methods can achieve very high accuracy, but they require the antennas to be calibrated, which can be a challenging task in real-life deployments. ML-based methods~\cite{ICC}\cite{gönültaş2021csibased} can relax the calibration requirement significantly but may require a significant amount of training data. 
E.g., ~\cite{ICC} achieves centimeter-level accuracy with a $100$-antenna massive MIMO system, by fusing $2$ fully connected neural networks (FCNNs). Each FCNN has $8$ hidden layers and takes instantaneous spatial covariance matrix and channel impulse responses as inputs, which provide information in both angular and delay domains. Given the advantage of not having to calibrate the antenna array and its relatively low computational complexity, \cite{ICC} is used for evaluating the radio-based localization system. The ML algorithm pipeline using the LuViRA dataset is shown in Fig. \ref{fig:block_diagram_radio} with examples from ``position A" in the \shortit{Grid110} trajectory.
\subsection{Audio System}

\begin{figure*}[tb!]
  \centering
    \includegraphics[width=1.0\linewidth,trim={0.2cm 0cm 0.1cm 0cm},clip]{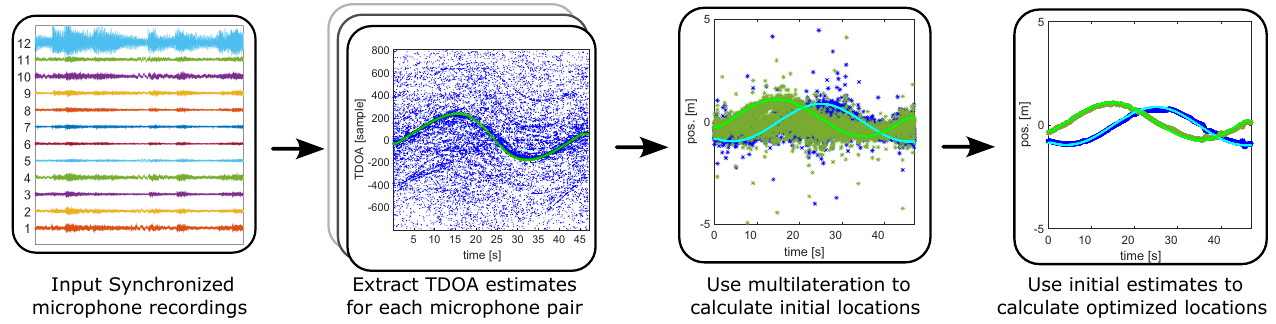}
  \caption{Selected audio-based localization algorithm pipeline for the RC2 trajectory.} 
  \label{fig:block_diagram_audio} 
\end{figure*}

Audio-based localization extracts distance information from audio recordings. One of the common setups is to have \textit{N} stationary microphones recording the sound from a single moving sound source for \textit{M} TDOA estimates. Under certain conditions, it is possible to simultaneously locate both the sound source and microphones. This is referred to as the structure from sound (SFS) problem~\cite{NIPS2005_a7a3d70c}. 
As in the radio case, the location can be estimated in several ways, such as by using strength, travel time, or the direction of arrival (using an array of microphones) of the sound signal. Time-based features such as TOA and TDOA, are widely used in solving the SFS problem~\cite{TDOA_1}\cite{zhayida2014automatic}. The TDOA feature is the basis for the SFS2 algorithm, which is an open-source localization algorithm that can estimate the sound source locations with or without prior knowledge of the microphone locations. 
Accuracy can be increased by the assumption that the sound source is tracked continuously during smooth motion, which enables using recent estimates as prior location information and temporal smoothness constraints.
The SFS2 algorithm consists of the following steps~\cite{9616051}: 
\begin{enumerate}
  \item Read synchronized microphone recordings as input.
  \item Extract initial TDOA estimates for each time instant and each pair of microphones by using Generalized Cross Correlation with Phase Transform (GCC-PHAT). 
  \item For each time instant, find and use the most suitable TDOA estimates to calculate an initial location of the microphones and the sound source (using minimal solvers, multilateration, and RANSAC to arbitrate among solutions).
  \item Use the (initial) location of the microphones and TDOA estimates in a robust multilateration method to calculate a final estimate of the sound source location. 
\end{enumerate}
Since GCC-PHAT uses the correlation between microphones to estimate TDOA between each pair, step $2$ could be modified to match towards a reference signal, e.g. a known chirp pattern. Such use of prior knowledge is not a part of the current implementation~\cite{larsson2021fast}. If the microphone locations are known, step $3$ is not required, and finding initial source locations is in step $4$. 
Under certain conditions, e.g., uncorrelated Gaussian noise stochasticity, the sound source location can be estimated with a few centimeters accuracy~\cite{larsson2021fast}. Thus, the SFS2 algorithm is selected as the benchmark method for audio data, using known microphone locations and smoothing. 
Fig. \ref{fig:block_diagram_audio} is an overview of how the algorithm is used with the LuViRA dataset.
\section{LUVIRA Dataset}
In this section, we describe the realistic measurement data used to validate the selected algorithms described in the previous section.
The LuViRA dataset~\cite{LuViRA} includes sensor data recorded by a camera, a massive MIMO antenna array, and $12$ microphones. The camera, a sound source (speaker), and a single-antenna radio transmitter are mounted to a robot so that only a planar movement is involved. During the measurements, a motion capture (mocap) system tracks these objects within $0.5$\mm accuracy with the help of markers, placed on the target objects. An overview of the measurement environment can be seen in Fig.~\ref{fig:Measurements}. The camera, speaker, and antenna are placed at different locations in the robot, as shown in Fig.~\ref{fig:Measurements}. 

The dataset is divided into two parts based on the movement pattern of the robot: ``grid" and ``random" data. The ``grid" data includes trajectories where the robot moves forward and backward, facing the same direction, in a measurement area of \qtyproduct[product-units=power]{4.2 x 2.5}{\m}. The ``random" data includes circular, diagonal, and rectangular trajectories, as well as arbitrary trajectories. Abbreviations used for each trajectory are given in Table \ref{tab:trajectory_abbreviations}. 
\begin{table}[tb!]
\caption{Random trajectory abbreviations}
\begin{center}
\begin{tabular}{|l|c|}
     \hline
     \textbf{Trajectory Name} & \textbf{Alias} \\  \hline
     Random rect1 & RR1 \\ \hline %
     Random rect2 & RR2 \\ \hline %
     Random circle1 & RC1 \\ \hline
     Random circle2 & RC2 \\ \hline
  \end{tabular}
  \quad
\begin{tabular}{|l|c|}
     \hline
     \textbf{Trajectory Name} & \textbf{Alias} \\  \hline
     Random manual people1 & RM1 \\ \hline
     Random manual people2 & RM2\\ \hline
     Random manual3 & RM3 \\  \hline
     Random manual4 & RM4 \\ \hline
     Random manual5 & RM5 \\ \hline
     Random manual people6 & RM6 \\ \hline
  \end{tabular}
  \end{center}
  \label{tab:trajectory_abbreviations}
\end{table}%
The vision, radio, and audio systems used to capture the dataset are briefly introduced next. 
\subsubsection{Vision System}
The Intel Realsense D435i depth camera is used for obtaining RGB images ($30$\,fps) and depth maps ($15$\,fps) from the environment. The \shortit{Grid110} and selected ``random" trajectories are used as input to RGB-D ORB-SLAM3 for evaluating localization accuracy. The selected trajectories include the ``Random manual" trajectories, where the robot makes sharp movements. In RM1, RM2, and RM6, people and objects are moving around. Automated evaluation tools~\cite{TUM_dataset} are used to calculate the absolute trajectory error. The evaluation tool aligns the ground truth labels and the estimated trajectory by solving for rigid transformation between trajectories. The field-of-view and frame rate of the provided depth maps and RGB images are aligned. 

\subsubsection{Radio System}
The measurement setup includes a single user equipment (UE) with one antenna and the LuMaMi testbed with $100$ antennas. Both the UE and the LuMaMi testbed have $20$\MHz bandwidth and $3.7$\GHz center frequency. In the LuViRA dataset, a cable is used to synchronize the UE and LuMaMi testbed to avoid possible signal drops. The ``grid" data, which consists of $75$ trajectories, is used to train and test the ML-based algorithm since localization estimates of the target object in the \say{random} trajectories are outside the measurement area defined by the dataset. In the reference paper of the selected radio-based ML algorithm~\cite{ICC}, both the training and testing data are randomly selected from all ``grid" trajectories, leading to training data that encompasses randomly chosen points from each trajectory.
Instead, we separate training and testing trajectories, so we can compare the localization accuracy of each sensor on a completely unused trajectory. To conduct a balanced comparison with localization methods based on other sensors, odd numbered trajectories (number $1$, $3$, $5$, and so on) are used to train the ML algorithm, whereas the accuracy of the algorithm is tested on the evenly numbered trajectories ($2$, $4$, etc.). 

\subsubsection{Audio System}
The audio data include recordings from $12$ microphones. One of these microphones is placed on the robot to record the sound directly from the speaker. This reference microphone is not used in the scope of this paper. The post-processed recordings are single-channel WAV-files and have a sample rate of $96$\kHz and $16$-bit resolution. 
The SFS2 algorithm is evaluated on the \shortit{Grid110}, RR1, and RR2 trajectories where the speaker plays a chirping sound and for the rest of the ``random" trajectories where the speaker plays different music clips.  
In these recordings, there is an audible fan noise from the LuMaMi testbed.
The signal-to-noise ratio (SNR) varies naturally as the sound source moves around, and microphone proximity to LuMaMi, but the signal is always audible. 
Considering that ground truth system calibration might fluctuate between trajectories, the microphone locations are assumed to be fixed for each trajectory separately. 


\section{Results and Discussion}

Table \ref{tab:table_compare} summarises the results when applying the selected localization algorithms to the realistic dataset. Some of the example trajectories can be seen in Fig.~\ref{fig:res_vision_grid}, \ref{fig:res_radio}, \ref{fig:res_audio_circle1}, \ref{fig:res_vision_random} and \ref{fig:res_audio_RM6ppl_XY}. In this section, different aspects of the results are discussed.
\begin{table}[tb!]
\caption{Localization results on selected trajectories}
\begin{center}
\begin{tabular}{|l|c|c|c|}
     \hline
      &  \textbf{Vision} & \textbf{Radio} & \textbf{Audio} \\  \hline
     Sensor & camera  & antenna array & microphones \\ \hline
     Localization & ORB - &  ML\cite{ICC} & SFS2\cite{9616051} \\
     algorithm &  SLAM3\cite{ORB-SLAM3} &  &  \\ \hline
     Estimation rate$^{\mathrm{a}}$   & $15$\,Hz & $100$\,Hz & $100$\,Hz \\ \hline 
     Mean error, Grid110& $6.5$\,cm & $14$\,cm  & $127$\,cm $^{\mathrm{b}}$ \\\hline
     SD error, Grid110 & $3.8$\,cm & $13$\,cm & $74$\,cm$^{\mathrm{b}}$ \\\hline
     Mean error, RC1 & $5.7$\,cm & N/A  & $15$\,cm \\\hline
     SD error, RC1 & $3.4$\,cm & N/A & $17$\,cm \\\hline
     Mean error, RM3 & $10$\,cm & N/A  & $6.4$\,cm \\\hline
     SD error, RM3 & $7.2$\,cm & N/A & $4.9$\,cm \\\hline
  \multicolumn{4}{l}{$^{\mathrm{a}}$How frequent location estimates are completed in this benchmark.}\\
  \multicolumn{4}{l}{$^{\mathrm{b}}$Trajectories using chirp audio.}  
  \end{tabular}
  \end{center}
  \label{tab:table_compare}
\end{table}
\subsection{Localization accuracy}
The accuracy of the localization algorithms is calculated based on the mean and SD of the Euclidean distance between the estimated locations and ground truth labels. The vision, radio, and audio systems calculate 6D, 3D, and 3D locations, respectively. Since the orientation is only available for the vision sensor, it is ignored in the accuracy calculations and the objects are located at a constant height over the floor, so the produced 3D locations are projected on the 2D plane to allow comparison between systems.
Prior knowledge of the environment affects the localization accuracy of all the sensors. ORB-SLAM3 assumes that the environment is previously unknown, whereas the ML algorithm requires extensive training to learn the radio features in the environment and the SFS2 algorithm assumes the microphone locations to be known. Please note that both ORB-SLAM3 and SFS2 are not fully deterministic algorithms, and exploring the effect of the random seed and its spread is out of the scope of this work.

\begin{figure}[tb!]
  \centering
  \includegraphics[width=0.9\linewidth]{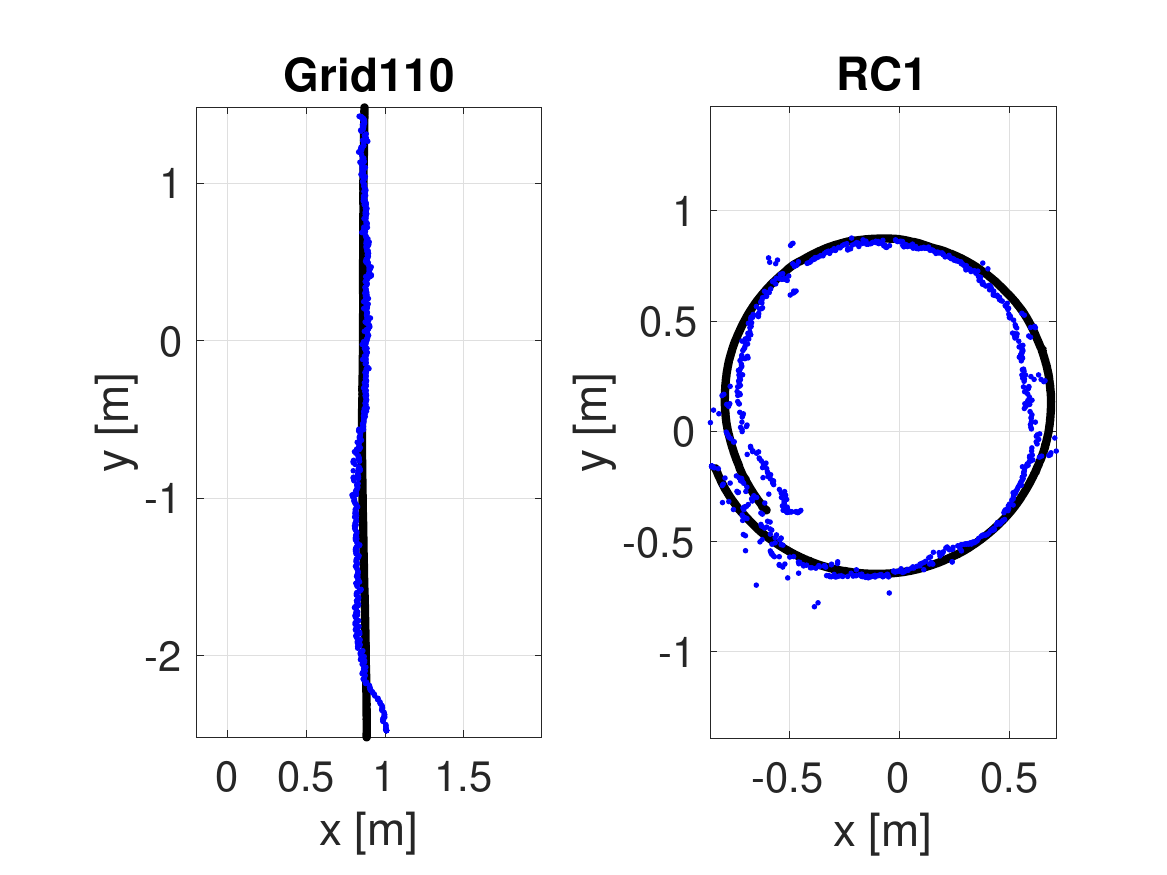}
  \caption{Ground truth (black) and the result of the ORB-SLAM3 (blue) for Grid110 and RC1.}
  \label{fig:res_vision_grid}
\end{figure}

The selected vision-based localization algorithm successfully localizes the target object in both the ``grid" and ``random" trajectories. To illustrate the results, we choose the \shortit{Grid110} and the RC1 trajectories and show the corresponding results in Fig.~\ref{fig:res_vision_grid}. The localization accuracy of the ORB-SLAM3 algorithm is given in Table~\ref{tab:table_vision_results} for $11$ trajectories. 

\begin{table}[htbp!]
\caption{Part of the results of the vision system}
\begin{center}
\begin{tabular}{|c|c|c|c|}
     \hline
     \textbf{Trajectory Name}&\textbf{Mean (cm)}&\textbf{SD (cm)}&  \textbf{Median (cm)} \\  \hline
     Grid110 & 5.6 & 2.5 & 5.4 \\ \hline
     RR1 & 9.4 & 4.0 & 8.6 \\ \hline %
     RR2 & 6.1 & 2.5 & 5.7 \\ \hline %
     RC1 & 5.7 & 3.4 & 4.6 \\ \hline
     RC2 & 5.5 & 3.7 & 4.4 \\ \hline
     RM1 & 29 & 12 & 24 \\ \hline
     RM2 & 56 & 30 & 50\\ \hline
     RM3 & 10 & 7.2 & 8.2 \\  \hline
     RM4 & 12 & 6.4 & 10 \\ \hline
     RM5 & 14 & 5.2 & 13 \\ \hline
     RM6 & 50 & 24 & 51\\ \hline
  \end{tabular}
  \label{tab:table_vision_results}
  \end{center}
\end{table}

\begin{figure}[tb!]
  \centering
  \includegraphics[trim=1.3cm 0.2cm 1.5cm 11cm, clip, width=1.0\linewidth]{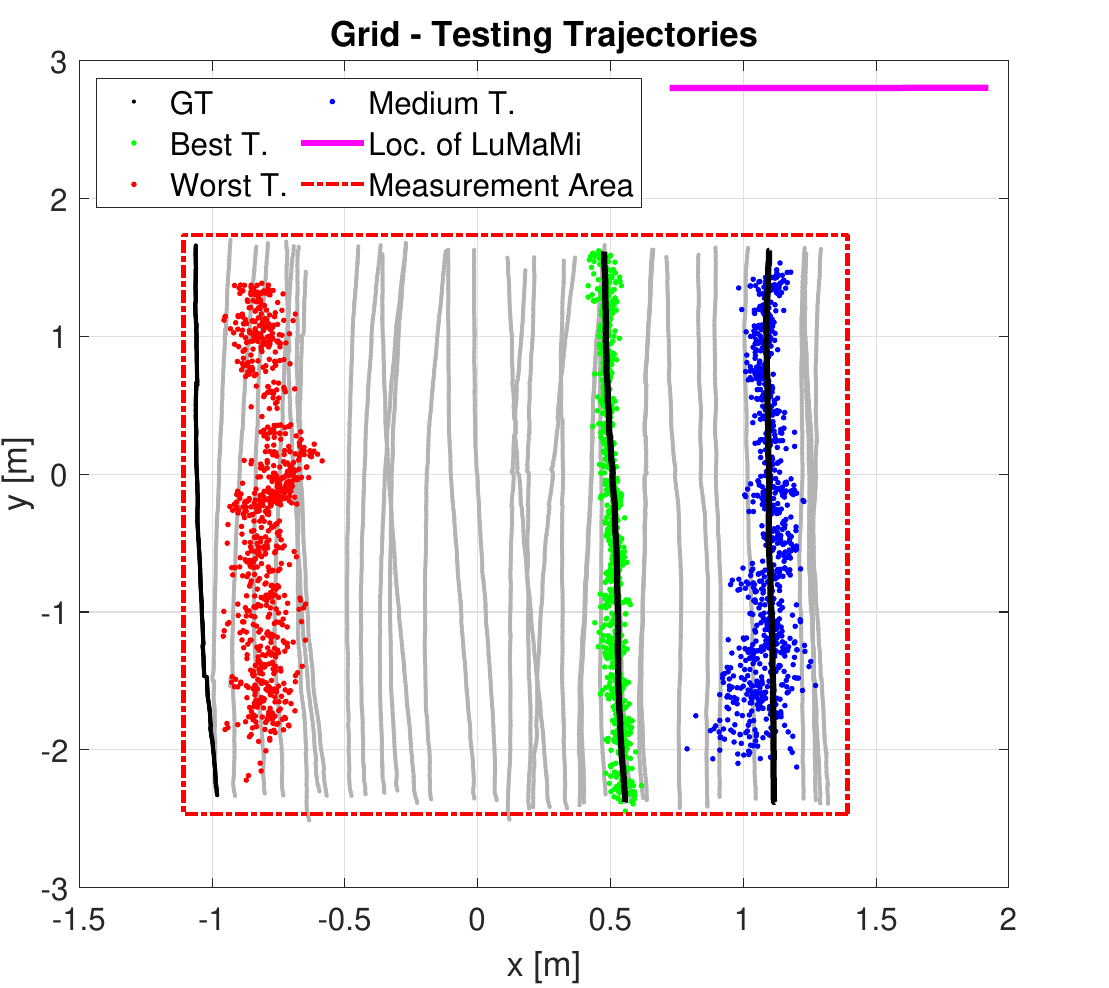}
  \caption{Overview of the ground truth (gray stripes) and the result of the ML algorithm for the testing trajectories with best (\shortit{Grid126}, in the middle), worst (\shortit{Grid172}, left side), and medium (\shortit{Grid110}, right side) localization accuracy. Their ground truth trajectories are shown in black. The measurement area (dashed red lines) and the antenna array (horizontal pink line) are illustrated too.} %
  \label{fig:res_radio}
\end{figure}

The radio localization algorithm results in an overall mean error of $13$\cm and SD of $12$\cm for the testing trajectories. The mean and SD of the trajectory with the performance closest to the overall mean error (medium estimated trajectory or \shortit{Grid110}), are presented in Table~\ref{tab:table_compare}. Fig.~\ref{fig:res_radio} gives an overview of all the testing trajectories and the estimation results of the best, worst, and medium trajectories based on the localization accuracy of the ML algorithm. The measurement area and the location of LuMaMi's antenna array are also given in red dashed lines and the pink line, respectively, as a reference. 
When training using the ``grid" trajectories and testing using the ``random" trajectories the estimated locations are not valid. 

The best and worst performances are achieved for \shortit{Grid126} and \shortit{Grid172} trajectories with a mean of $5$\cm and $35$\cm and SD of $4$\cm and $11$\cm, respectively. In Fig.~\ref{fig:res_radio}, the test trajectory furthest away from the antenna array performs the worst while the performance declines when the robot moves away from the antenna array (in the y-direction) for the medium estimated trajectory. This can be explained by the increased distance from the base station and the number of training samples available~\cite{guoda_journal}. On the other hand, the localization accuracy could be improved by adding an extended Kalman filter (EKF), particle filters, or other methods that apply tracking or smoothing~\cite{rovio}\cite{Kowalski_smoothing}. Such methods are used in both ORB-SLAM3 and SFS2 algorithms to correct errors based on the prior location of the tracked object. 

\begin{figure}[tb!]
  \centering
\begin{minipage}{0.5\linewidth}
  \centering
  \includegraphics[width=1.0\linewidth]{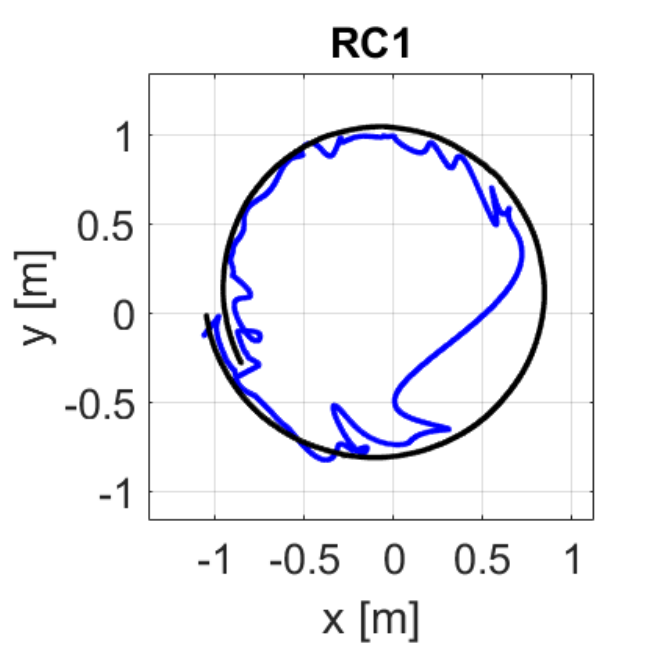}
\end{minipage}%
\begin{minipage}{.5\linewidth}
  \centering
  \includegraphics[width=0.95\linewidth]{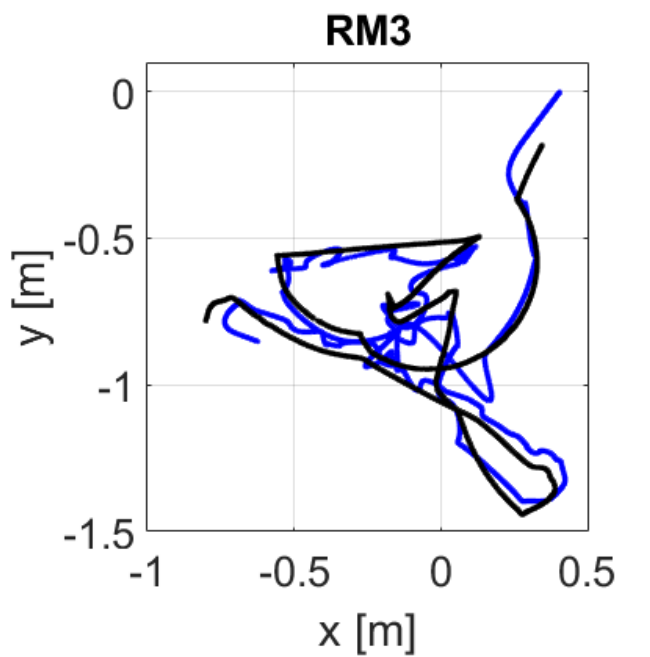}
\end{minipage}
  \caption{Ground truth (black) and the result (blue) of SFS2 algorithm given known microphone locations for RC1 and RM3 trajectories.}
  \label{fig:res_audio_circle1}
\end{figure}

A part of the results for the SFS2 algorithm is given in Table~\ref{tab:table_audio_results}. The median localization error values are also presented in the table to illustrate the distribution of the results. E.g., the RC1 trajectory shown in Fig.~\ref{fig:res_audio_circle1}, has $15$\cm, $17$\cm mean and SD error, whereas the median is only $7.2$\cm, indicating the results are not normally distributed.
As seen in Table~\ref{tab:table_audio_results}, trajectories with chirp-based audio signal results in more than a meter localization error, while the music-based trajectories, such as RM3, can achieve up to $6.4$\cm mean localization accuracy. The differences between chirp and music based trajectories will be discussed in the next subsection. 

\begin{table}[tb!]
\caption{Part of the results of the audio system}
\begin{center}
\begin{tabular}{|c|c|c|c|}
     \hline
     \textbf{Trajectory Name}&\textbf{Mean (cm)}&\textbf{SD (cm)}&\textbf{Median (cm)}\\  \hline %
Grid110$^{\mathrm{a}}$ & 127 & 74 & 116 \\ \hline %
RR1$^{\mathrm{a}}$ & 152 & 73 & 130 \\ \hline %
RR2$^{\mathrm{a}}$ & 167 & 158  & 111\\ \hline %
RC1 & 15 & 17 & 7.2 \\ \hline %
RC2 & 9.6 & 3.1 & 9.3 \\ \hline %
RM1 & 31 & 44  & 9.3 \\ \hline %
RM2 & 28 & 38 & 16 \\ \hline %
RM3 & 6.4 & 4.9 & 5.2 \\ \hline
RM4 & 21 & 28 & 8.7 \\ \hline %
RM5 & 49 & 95 & 8.7 \\ \hline %
RM6 & 34 & 53 & 16 \\ \hline
\multicolumn{4}{l}{$^{\mathrm{a}}$Trajectories using chirp audio.}
  \end{tabular}
  \label{tab:table_audio_results}
  \end{center}
\end{table}
\subsection{Sensitivity to dynamic environment}
\begin{figure}[tb!]
  \centering
  \includegraphics[width=1.0\linewidth]{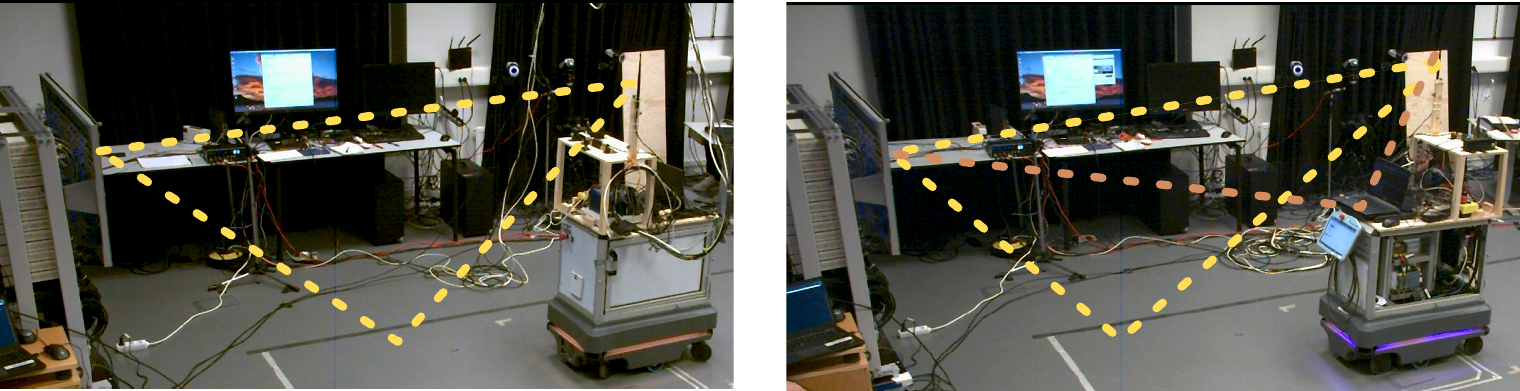}
  \caption{Conceptual illustration of the change in the propagation environment (dashed, yellow) for different robot orientations in \shortit{Grid122} (left image) and RC1 trajectories (right image). The difference is highlighted with orange color.}
  \label{fig:radio_multipath}
\end{figure}

Changes in the propagation environment affect radio-based localization algorithms in terms of reliability. As mentioned above, the orientation of the robot changes while capturing data for the ``grid" and ``random" trajectories. For example, the location of the robot is always in between the UE and LuMaMi for the ``grid" trajectories while for the random trajectories, it can be on the side, changing the multipath components in the environment, which results in different radio channel responses for a given location. This change is illustrated in Fig. \ref{fig:radio_multipath}. 

\begin{figure*}[tb!]
  \centering
  \includegraphics[trim=5.9cm 0 13.0cm 73cm, clip,width=0.9\linewidth]{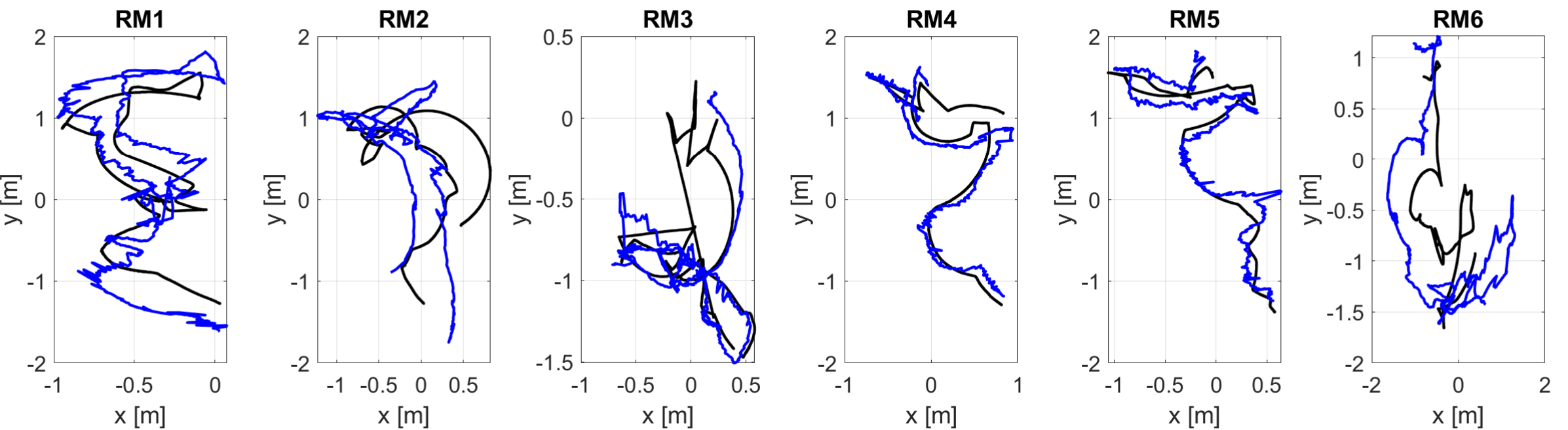}
  \caption{Ground truth (black) and the result of the ORB-SLAM3 (blue) for the ``Random manual" (RM) trajectories. The comparably worse tracking of RM1, RM2, and RM6 are the trajectories with non-static features or dynamic environment where people are moving around.}
  \label{fig:res_vision_random}
\end{figure*}

On the vision system side, having static objects with distinct features in the environment is an important requirement for achieving high accuracy with the ORB-SLAM3 algorithm. If a room is empty and the walls are just white, the algorithm cannot extract features (since there are none) from the given frames and the accuracy of the algorithm drops significantly \cite{low_texture2}. Thus, the measurement environment is decorated using posters, a chess board, etc, as seen in Fig. \ref{fig:Measurements} and \ref{fig:block_diagram_vision}. Moreover, if the camera is moving at a very high speed with many rotations, the frame rate of the camera is not sufficient. E.g., the RM3, RM4, and RM5 trajectories in Table \ref{tab:table_vision_results}, have sharp turns which result in a slight increase of the localization error compared to other trajectories that are not labeled ``manual". In contrast to the other sensors, increasing the frame rate to more than $60$\,fps is challenging due to the associated bandwidth, memory, and power requirements. The RM1, RM2, and RM6 trajectories also have non-static features which result in a sharp decline in the accuracy. The results of all the \say{Random manual} trajectories are shown in Fig. \ref{fig:res_vision_random}.

\begin{figure}[tb!]
\begin{minipage}{0.6\linewidth}
  \centering
  \includegraphics[width=0.9\linewidth,trim={0cm 0cm 0.0cm 0cm},clip]{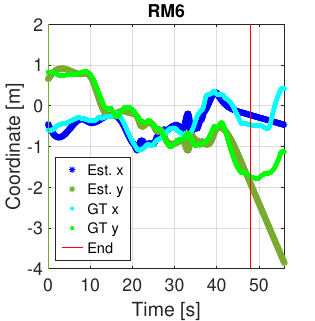}
\end{minipage}%
\begin{minipage}{0.4\linewidth}
  \centering
  \includegraphics[width=1\linewidth,trim={0cm 0cm 0.1cm 0cm},clip]{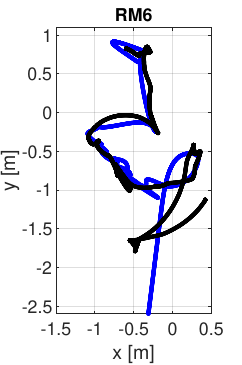}
\end{minipage}
  \caption{The left image shows, over time, the estimated X (dark blue) and Y (dark green) coordinates, and ground truth X (light blue) and Y (light green). The vertical red line shows where the audio signal ends. The right image shows the ground truth (black) and the result (blue).} 
  \label{fig:res_audio_RM6ppl_XY}
\end{figure}

For the audio system, the direct path between the speakers and the microphones is obstructed for RM1, RM2, and RM6, which impacts the extraction of the TDOA features. Even though this impact is expected to be minimal due to the distribution of microphones, the results for these trajectories are worse than the RM3, RM4, and RM5 trajectories. Another explanation can be the other sound sources such as footsteps and occasional speech in the audio recordings which only exist in RM1, RM2, and RM6 trajectories. Moreover, RM6 has the music fading and stopping ahead of time, which results in detrimental signal loss, as can be seen in \autoref{fig:res_audio_RM6ppl_XY}. 


\subsection{Sensitivity to Signal-to-Noise Ratio} 

The accuracy of vision-based localization algorithms is reduced drastically when there is little to no light (night or darkness), interference with other strong light sources such as the sun, or when the camera is blocked by other objects~\cite{slam_light}. To simulate low-light conditions, where the thermal noise in the camera has a greater effect on the image, a zero-mean, Gaussian white noise is added to the RGB images obtained by the camera. Further investigation on adding noise to the depth maps is left as future work. 
These results are compared to the radio system, where the same noise level is added to both training and testing data with SNR of \qty{10}{\deci\bel} and \qty{20}{\deci\bel}. The noise is applied before any processing is made, including the alignment of images and training of the neural network. The ORB-SLAM3 algorithm is highly affected by the low SNR even when the noise is only added to the RGB images. The radio-based localization algorithm is more robust against low SNR resulting from added white Gaussian noise compared to the vision system. One possible explanation for the difference between these sensors is the redundancy in the 100-antenna matrix and 100-subcarrier channel state information provided by LuMaMi testbed. Another explanation is how these algorithms process the given inputs. The ORB-SLAM3 algorithm processes its inputs (frames) as a whole and extracts information (such as ORB features) while the ML-based algorithm learns possible changes in SNR and uses the strength and phase of the radio signal directly. 

In many realistic audio-based localization systems, sound sources such as speech, do not transmit continuous signals and may have a limited frequency range. The SFS2 algorithm does not use prior knowledge about the sound signal, instead, the correlation between the microphone pairs is used as a way to separate signal from noise. This works better for wideband signals, such as music, than narrowband signals where a larger share of the spectrum can be dominated by correlated or uncorrelated noise. 

\begin{table}[htbp!]
\caption{Results of adding Gaussian noise to vision and radio sensors for Grid110 trajectory}
\begin{center}
\begin{tabular}{|l|c|c|c|}
     \hline
     \textbf{Sensor} & \textbf{SNR (dB)} & \textbf{Mean (cm)} &  \textbf{SD (cm)} \\  \hline
     \multirow{3}{*}{Vision} & N/A$^{\mathrm{a}}$ & 6.5 & 3.8 \\ \cline{2-4}
     & 10 & 89 & 48 \\\cline{2-4}
     & 20 & 22 & 10 \\\hline
     \multirow{3}{*}{Radio} & N/A$^{\mathrm{a}}$ & 14 & 13 \\ \cline{2-4}
      & 10 & 15 & 14 \\ \cline{2-4}
      & 20 & 14 & 13 \\ \hline
     
  \multicolumn{4}{l}{$^{\mathrm{a}}$Baseline (no added noise)}
  \end{tabular}
  \label{tab:table_vision_noise}
  \end{center}
\end{table}

\begin{figure}[tb!]
\begin{minipage}{0.5\linewidth}
  \centering
  \includegraphics[width=0.9\linewidth]{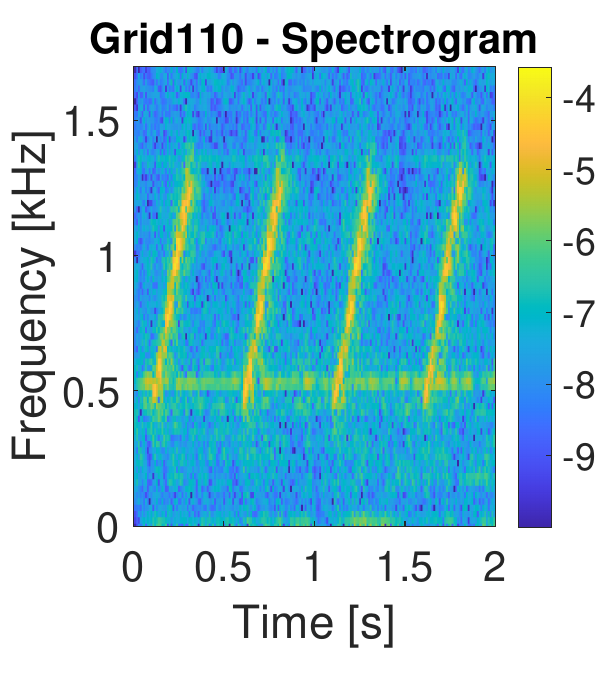}
\end{minipage}%
\begin{minipage}{.5\linewidth}
  \centering
  \includegraphics[width=1.0\linewidth]{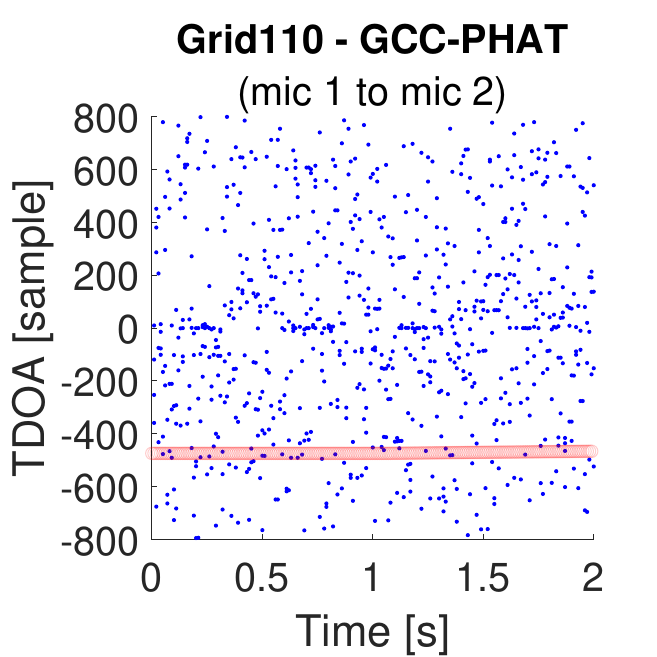}
\end{minipage}
  \caption{The left image shows the power spectrogram of the signal from microphone 1. The right image displays the estimated TDOA (blue dots) and the expected TDOA based on the ground truth (red area).  
  }
  \label{fig:chirp_spectrum}
  \label{fig:audio_peaks_grid10}
\end{figure}

Table \ref{tab:table_audio_results} shows that the selected audio system fails for chirp-based trajectories. Fig. \ref{fig:audio_peaks_grid10} shows an example of extracted TDOA estimates for \shortit{Grid110}, with a ground truth overlay. The periodic clusters at zero TDOA may correspond to the silent time periods, but there is no pattern along the ground truth labels. Since the algorithm has no access to the ground truth, it randomly bases most estimates on the noise. This can be compared to a successful localization example, such as \autoref{fig:block_diagram_audio} where step 2 extracts well-estimated TDOA features. From the spectrogram analysis of the chirp audio given in \autoref{fig:chirp_spectrum}, the signal is estimated to be a sine wave rising from \qtyrange{400}{1400}{\hertz} during \val[\ms]{200}. It repeats twice per second, which leaves $60$\,\% of the time, silent where the audio system relies on interpolation. In a reverberant room, the challenge is even greater, as echoes may be interpreted as signals and result in misleading estimates. The \autoref{fig:radio_multipath} illustration of multi-path arrival also in this case. The many reflection surfaces around the speaker result in an ever-present source of correlated noise. 

\subsection{Calibration requirements} 
Each of the discussed sensors requires different intrinsic and extrinsic calibration and time synchronization of different components in the system. 
Cameras used in the ORB-SLAM3 algorithm need to be calibrated once before the camera starts recording. This calibration configures the intrinsic parameters, such as focal length and distortion of the lens, which are required in the selected algorithm.
The LuMaMi testbed performs the reciprocity calibration automatically. However, the characterization of the antenna array elements is unknown, which is not an obstacle for ML-based localization algorithms as long as coherency is maintained. 
Moreover, a cable is used to synchronize the UE and LuMaMi testbed, which is usually not feasible in real-life scenarios.  
For the audio system, the synchronization between microphones needs to be very accurate, thus they are connected to a single sound card via cables. Moreover, the changes in humidity and temperature of the room could invalidate the calibration and have an impact on all the sensors \cite{temparature_camera}. The SFS2 algorithm is directly affected by the change since it uses the speed of sound to calculate the distance. According to the description of the LuViRA dataset, the temperature changed from \qty{22}{\celsius} to \qty{28}{\celsius} which corresponds to a $3.6$\,\unit{\metre\per\second} change in the speed of sound. For reference, the speed of sound at the initial room temperature is $344$\,\unit{\metre\per\second}. 
Thus, a $1$\% error in the speed of sound will have a below-centimeter impact in the given measurement area of \qtyproduct[product-units=power]{4.2 x 2.5}{\m}. In some cases, the smoothing forces the estimates to stray toward an outlier that is perceived as confident. There is a bias to give estimates around $(-0.5,-0.5)$ which might be related to the position of the fans. 
Trajectories that pass near this point might achieve a better accuracy, partly because these outliers happen to result in a low error and partly because the smoothing can correct some of them. 

The processed ground truth follows virtual points based on the markers. In reality, these points may have an offset against the target objects, e.g. the speaker. For the audio system, \autoref{fig:res_audio_circle1} may indicate that the ground truth labels are further away than the estimated trajectory. However, no bias is confirmed, given the large SD. 
A similar bias could affect the other sensor estimates (e.g. \autoref{fig:res_vision_grid}), although with an unrelated bias since they track different points. These systems may compensate for such bias. The radio system is trained to predict the ground truth (including any consistent bias) and the vision system gives relative locations that are aligned to the ground truth. The audio system estimates absolute locations (relative to the microphones) without any reference to ground truth labels. 

\section{Conclusion \& Further Work} 
In this paper, we evaluated vision, radio, and audio based localization methods, using state-of-the-art algorithms for each sensor type and the LuViRA dataset. Aspects including localization accuracy, reliability, calibration requirements, and potential system complexity are discussed to analyze the advantages and limitations of using different sensors for indoor localization tasks. 
To sum up, each sensor has its advantages and limitations. It is challenging to find a sensor that is suitable for all situations. The vision-based localization algorithm can localize all the trajectories in the LuViRA dataset while the audio system has lower accuracy in trajectories with a chirp signal as the sound source and radio-based localization algorithm cannot localize the ``random" trajectories. However, when objects or people are moving around, the audio-based localization algorithm reaches higher accuracy and may be the most robust (as long as there is no additional sound in the environment from the movement). The current implementation of the radio-based localization algorithm cannot successfully localize the ``random" data in the dataset but, it is more robust against low SNR conditions. 
Pursuing algorithms that work well on audio and radio for both the \say{grid} and the \say{random} trajectories are left as future work. To leverage the strengths of each medium and overcome the weaknesses, a fusing strategy that jointly processes the information and increases algorithm robustness should be developed. 

\section*{Acknowledgment}

The authors would like to thank Henrik Garde from Lund University Humanities Lab, Alexander Dürr and Volker Kruger from the Department of Computer Science, Martin Larsson from the Centre for Mathematical Sciences, Anders Robertsson from the Department of Automatic Control, and Michiel Sandra, Sirvan Abdollah Poor, Jesús Rodríguez Sánchez and Sara Willhammar from the Department of Electrical and Information Technology at Lund University for providing resources, technical support, and assistance in our project. 
\section*{References}
\bibliographystyle{IEEEtran}
\bibliography{main}
%
\section*{} 
\vskip -6\baselineskip plus -1fil
\begin{IEEEbiography}[{\includegraphics[width=1in,height=1.25in,clip,keepaspectratio]{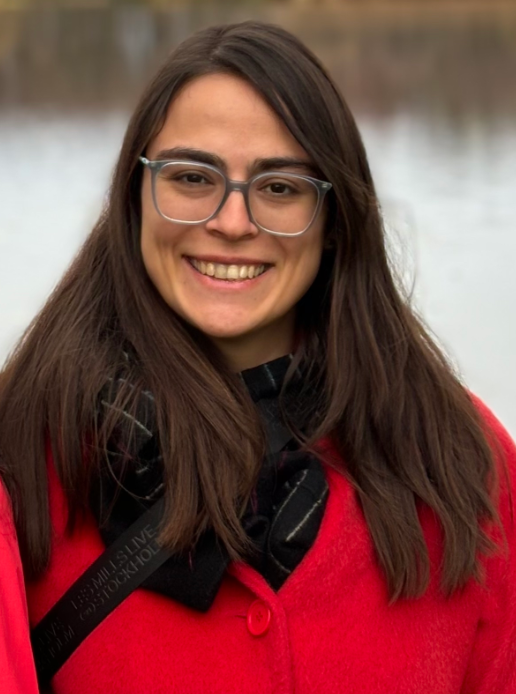}}]
 {Ilayda Yaman}~(Student Member, IEEE) 
completed her bachelor’s degree at Istanbul Technical University in 2018 and her master’s degree in Embedded Electronics Engineering at Lund University in 2020. During her master's degree, she received the LU Global Scholarship. Currently, she is a Ph.D. student at Lund University (Main Supervisor: Liang Liu). Her current research area is low-power ML hardware for vision and radio based systems.
\end{IEEEbiography}
\vskip -1\baselineskip plus -1fil
\begin{IEEEbiography}[{\includegraphics[width=1in,height=1.25in,clip,keepaspectratio]{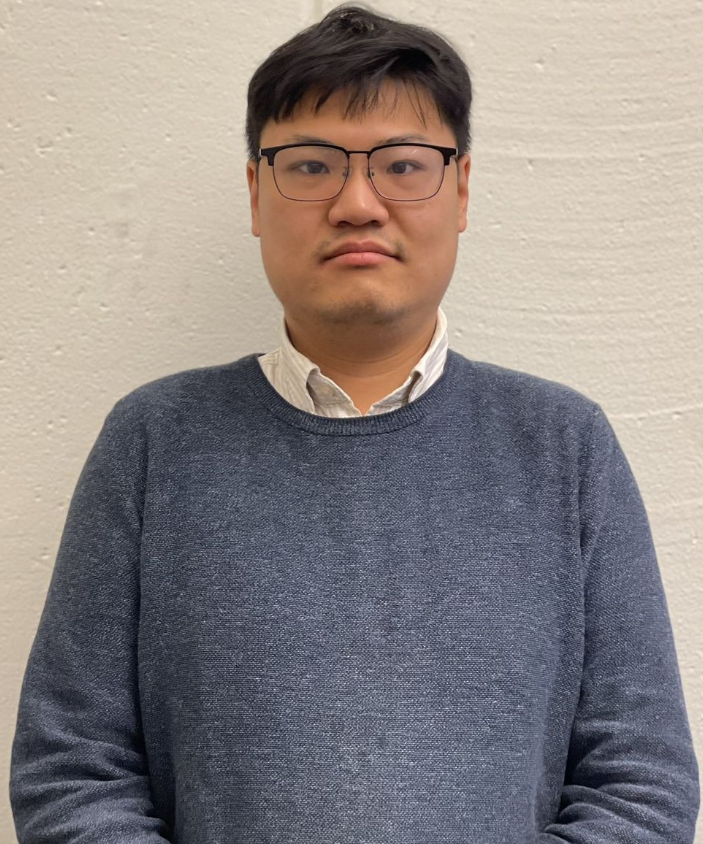}}]
{Guoda Tian}~(Student Member, IEEE) finished his bachelor’s degree in automation and control engineering at Northeastern University in 2016 and his master’s degree in wireless communication at Lund University in 2018 where he received the LU Global Scholarship. Currently, he is a Ph.D. student at Lund University  (Main Supervisor: Fredrik Tufvesson). His research topic is applying machine learning to wireless localization, sensing, and communication systems.
 \end{IEEEbiography}
\vskip -1\baselineskip plus -1fil
\begin{IEEEbiography}[{\includegraphics[width=1in,height=1.25in,clip,keepaspectratio]{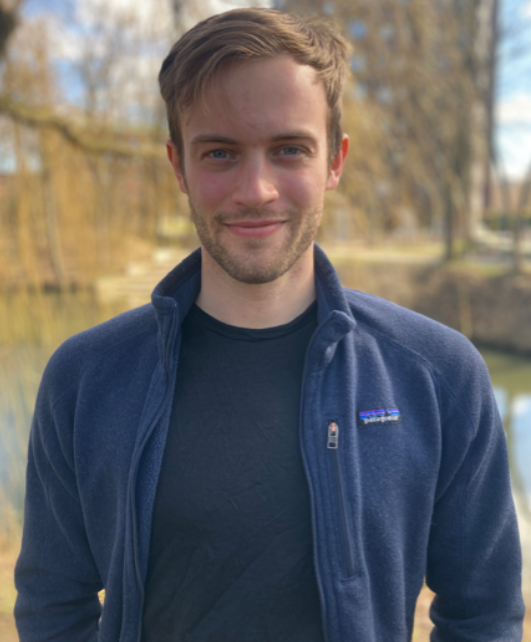}}]
 {Erik Tegler}~(Student Member, IEEE) completed his master’s degree in Engineering Physics at Lund University in 2021. He is currently a third-year Ph.D. student at Lund University under the supervision of Kalle Åström, Magnus Oskarsson, Bo Bernhardsson and Fredrik Tufvesson. His current research area is machine learning and its application in localization systems.
\end{IEEEbiography}
\vskip -1\baselineskip plus -1fil
\begin{IEEEbiography}[{\includegraphics[width=1in,height=1.25in,clip,keepaspectratio]{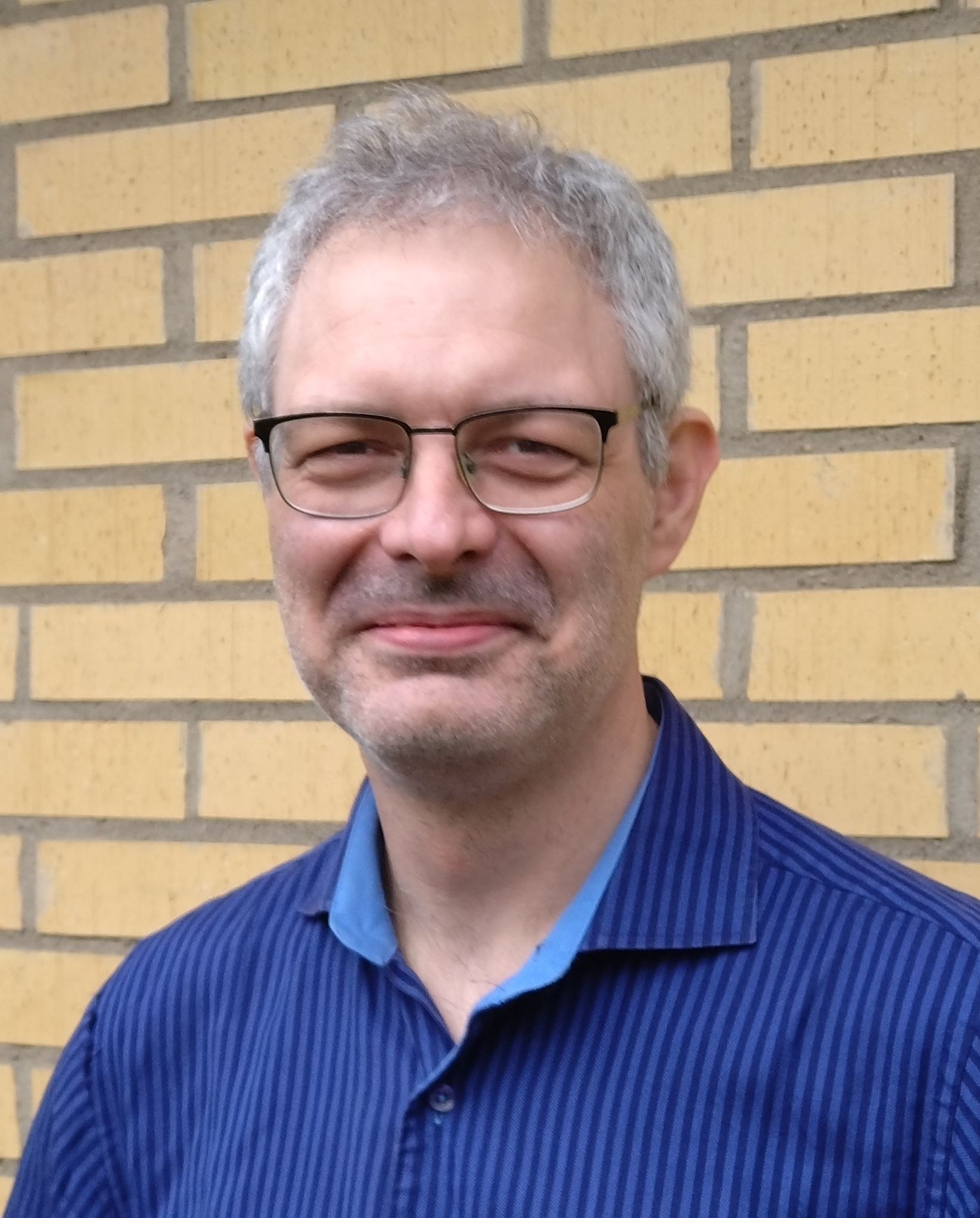}}]
{Jens Gulin}~(Student Member, IEEE) received his Master's degree in Computer Science and Engineering from Lund University in 2001. After many years in the mobile communication industry, the WASP initiative gave an opportunity to reconnect with the university to pursue a Ph.D (Main Supervisor: Kalle Åström) while employed by Sony in Lund. His research interest is systems able to learn from little data, here specifically for audio localization.
\end{IEEEbiography}
\vskip -1\baselineskip plus -1fil
\begin{IEEEbiography}[{\includegraphics[width=1in,height=1.25in,clip,keepaspectratio]{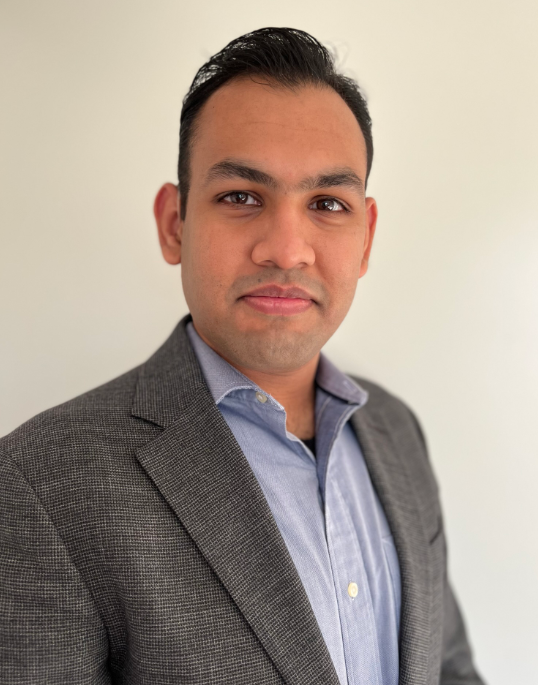}}]
 {Nikhil Challa}~(Student Member, IEEE) 
completed his bachelor’s degree at Visvesvaraya National Institute of Technology in 2008 and his master’s degree in Electrical Engineering at Virginia Tech in 2011. Afterwards, he worked at Qualcomm for 10 years. In 2023, he completed another master's program in Machine Learning, Systems and Control at Lund University. He is currently an ML engineer at Scania. 
\end{IEEEbiography}
\vskip -1\baselineskip plus -1fil
\begin{IEEEbiography}[{\includegraphics[width=1in,height=1.25in,clip,keepaspectratio]{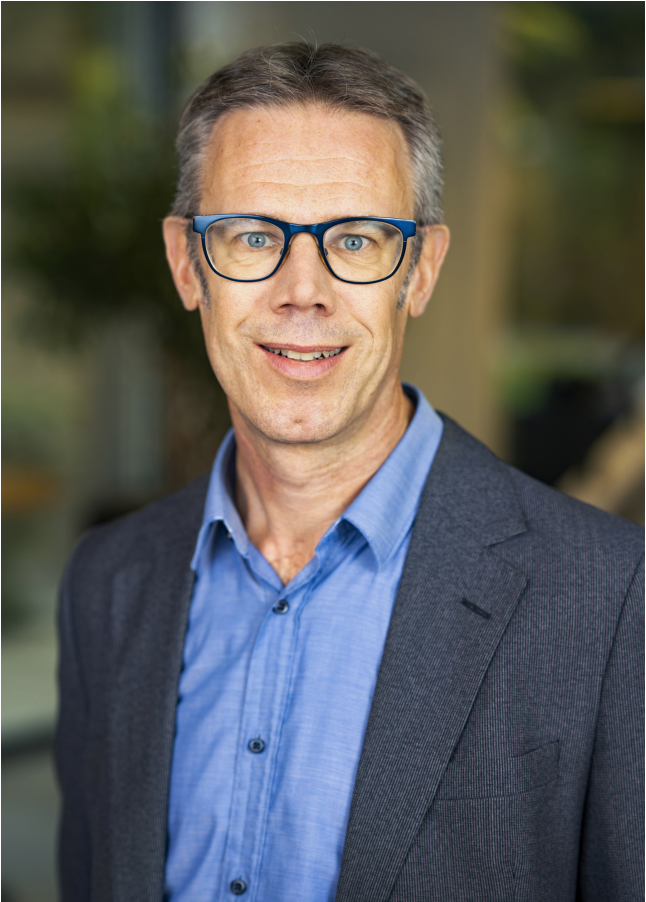}}] {Fredrik Tufvesson} ~(Fellow, IEEE) received the Ph.D. degree from Lund University in 2000. He is a professor of radio systems at the Department of Electrical and Information Technology, Lund University. His main research interest is the interplay between the radio channel and the rest of the communication system with various applications in 5G/B5G systems, such as massive MIMO, mmWave, vehicular communication, and radio-based positioning. 
\end{IEEEbiography}
\vskip -1\baselineskip plus -1fil
\begin{IEEEbiography}[{\includegraphics[width=1in,height=1.25in,clip,keepaspectratio]{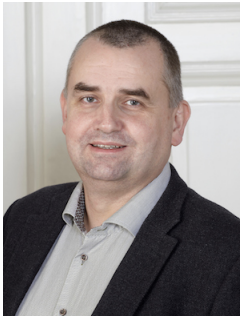}}] {Ove Edfors}~(Senior Member, IEEE) is Professor of Radio Systems at the Department of Electrical and Information Technology, Lund University, Sweden. His research interests include statistical signal processing and low complexity algorithms with applications in wireless communications. In the context of Massive MIMO and large intelligent surfaces, his main research focus is on how realistic propagation characteristics influence system performance and base-band processing complexity
 \end{IEEEbiography}
\vskip -1\baselineskip plus -1fil
 \begin{IEEEbiography}[{\includegraphics[width=1in,height=1.25in,clip,keepaspectratio]{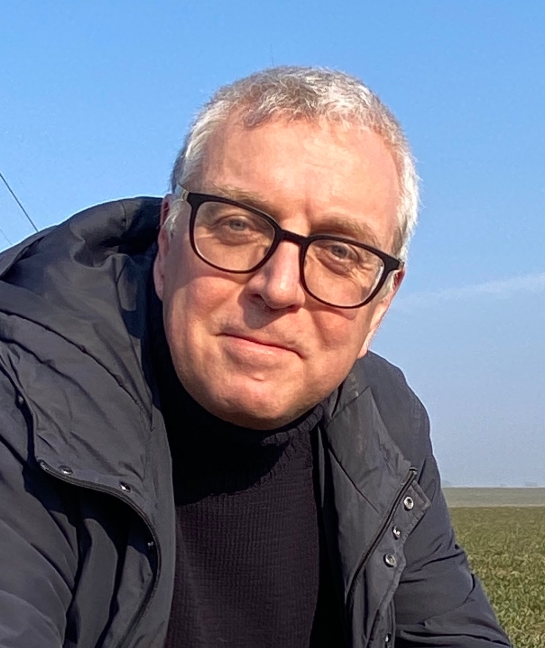}}] {Kalle Åström}~(Senior Member, IEEE) received his B.Sc in Mathematics in 1990, M.Sc. degree in Engineering Physics in 1991 and Ph.D. in Mathematics in 1996 from Lund University, Sweden. His thesis was awarded Best Nordic Ph.D. Thesis in pattern recognition and image analysis 1995-1996 at the Scandinavian Conference in Image Analysis, 1997. Currently, he is a professor at the Centre for Mathematical Sciences, LU. 
 \end{IEEEbiography}
\vskip -1\baselineskip plus -1fil
 \begin{IEEEbiography}[{\includegraphics[width=1in,height=1.25in,clip,keepaspectratio]{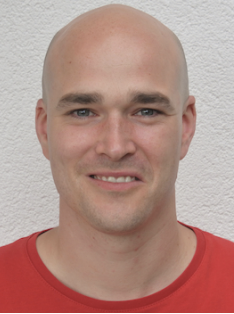}}] {Steffen Malkowsky} received his bachelor's degree in electrical engineering and information technology from Pforzheim University in 2011, his master's degree in electronic design from Lund University, in 2013, and his Ph.D. degree from Lund University, in 2019. Between 2019 and 2023, he worked as a Post-Doctoral Researcher at Lund University. Currently, he is the founder and CTO of Inceptron. 
 \end{IEEEbiography}
\vskip -1\baselineskip plus -1fil
\begin{IEEEbiography}[{\includegraphics[width=1in,height=1.25in,clip,keepaspectratio]{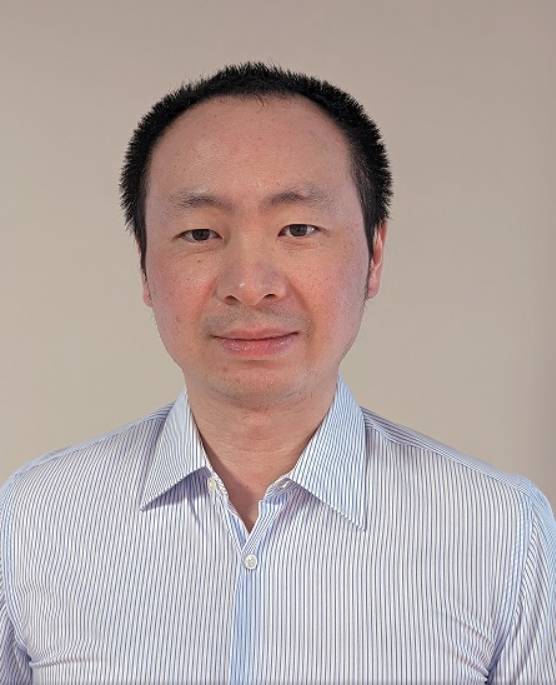}}] {Liang Liu}~(Member, IEEE) received his Ph.D. degree from Fudan University in 2010. He joined Lund University as post-doc. Since 2016, he has been an Associate Professor at Lund University. His research interests include wireless systems and digital integrated circuits design. He is a member of the Technical Committee of VLSI Systems and Applications and CAS for Communications of the IEEE Circuit and Systems Society.
 \end{IEEEbiography}
\end{document}